\documentclass[conference]{IEEEtran}
\IEEEoverridecommandlockouts
\usepackage{cite}
\usepackage{amsmath,amssymb,amsfonts}
\usepackage{graphicx}
\usepackage{textcomp}
\usepackage{xcolor}
\usepackage{makecell}
\usepackage{pifont}
\usepackage{xcolor}
\usepackage{array}
\usepackage{hyperref}
\usepackage{subcaption}
\usepackage{caption}
\usepackage{booktabs}
\usepackage{longtable}
\usepackage{amsmath}
\usepackage{algpseudocode}
\usepackage{hyperref}
\usepackage{amsfonts}
\usepackage{multicol}
\usepackage{arydshln}
\usepackage{hyperref}
\usepackage{tabularx}
\usepackage{pifont}
\usepackage{graphicx}
\usepackage{subcaption}
\usepackage{booktabs}
\usepackage{caption}
\usepackage{array}
\usepackage{xcolor}
\usepackage{colortbl}
\usepackage{float}
\usepackage{pgfplots}
\usepackage{pgfplotstable}
\usepackage{tikz}
\usetikzlibrary{patterns}
\usepackage{subcaption}
\usepackage{booktabs}
\usepackage{array}
\usepackage{multirow}
\usepackage{xcolor}
\pgfplotsset{compat=1.18}
\usepackage[table]{xcolor}  
\usepackage{booktabs,tabularx,makecell,multirow,subcaption}
\newcolumntype{Y}{>{\centering\arraybackslash}X}  % centred, stretchable

\usepackage{tikz}
\usetikzlibrary{positioning,arrows,calc}
\usepackage{textcomp}               % for \textdegree
\usepackage{tikz}
\usetikzlibrary{arrows.meta,positioning,trees}

\usepackage{verbatim}
\usepackage{xcolor}                      % node fill colors (gray!10, blue!8)

\usepackage{amsmath,amssymb,amsfonts}
\def\BibTeX{{\rm B\kern-.05em{\sc i\kern-.025em b}\kern-.08em
    T\kern-.1667em\lower.7ex\hbox{E}\kern-.125emX}}
\begin{document}

%\title{KORAL: Knowledge-Orchestrated Reasoning Using LLMs for Descriptive, Predictive, Prescriptive and What-If SSD Data Analytics}
% \title{From Fragmented Telemetry to Prescriptive and What-If SSD Insights via LLM + KG}

% \title{From Fragmented Telemetry to Operational SSD Insights via LLM + KG}
\title{KORAL: Knowledge Graph Guided LLM Reasoning for SSD Operational Analysis

\thanks{This paper has been accepted for publication in the Proceedings of the 40th IEEE International Parallel and Distributed Processing Symposium (IPDPS 2026), New Orleans, USA, May 25--29, 2026.}%
}


% --- Compact, “ACL-like” single-block authors (works in IEEEtran) ---
\author{
\IEEEauthorblockN{
Mayur Akewar\IEEEauthorrefmark{2},
Sandeep Madireddy\IEEEauthorrefmark{3},
Dongsheng Luo\IEEEauthorrefmark{2},
Janki Bhimani\IEEEauthorrefmark{2}
}
\IEEEauthorblockA{\IEEEauthorrefmark{2}\textit{Florida International University, Miami, FL, USA}}
\IEEEauthorblockA{\IEEEauthorrefmark{3}\textit{Argonne National Laboratory, Lemont, IL, USA}}
\IEEEauthorblockA{\texttt{makew001@fiu.edu, smadireddy@anl.gov, dluo@fiu.edu, jbhimani@fiu.edu}}
}

\maketitle

\begin{abstract}

Solid State Drives (SSDs) are critical to datacenters, consumer platforms, and mission-critical systems. Yet diagnosing their performance and reliability is difficult because data are fragmented and time-disjoint, and existing methods demand large datasets and expert input while offering only limited insights. Degradation arises not only from shifting workloads and evolving architectures but also from environmental factors such as temperature, humidity, and vibration. We present KORAL, a knowledge driven reasoning framework that integrates Large Language Models (LLMs) with a structured Knowledge Graph (KG) to generate insights into SSD operations. Unlike traditional approaches that require extensive expert input and large datasets, KORAL generates a Data KG from fragmented telemetry and integrates a Literature KG that already organizes knowledge from literature, reports, and traces. This turns unstructured sources into a queryable graph and telemetry into structured knowledge, and both the Graphs guide the LLM to deliver evidence-based, explainable analysis aligned with the domain vocabulary and constraints. Evaluation using real production traces shows that the KORAL delivers expert-level diagnosis and recommendations, supported by grounded explanations that improve reasoning transparency, guide operator decisions, reduce manual effort, and provide actionable insights to improve service quality. 
To our knowledge, this is the first end-to-end system that combines LLMs and KGs for full-spectrum SSD reasoning including Descriptive, Predictive, Prescriptive, and What-if analysis. We release the generated SSD-specific KG to advance reproducible research in knowledge-based storage system analysis. \textit{GitHub Repository: \href{https://github.com/Damrl-lab/KORAL}{Damrl-lab/KORAL}}

\end{abstract}

\begin{IEEEkeywords}
SSD, LLM, Ontology, Knowledge Graph
\end{IEEEkeywords}
\section{Introduction and Motivation}
\label{sec:intro}

Solid State Drives (SSDs) are the standard choice for fast and energy efficient storage in large scale datacenters \cite{b3}, personal computers \cite{b4}, and safety critical systems in aerospace and defense \cite{radiation}. Their low latency, high throughput, and compact size raise the quality of service (QoS) for data intensive work such as web search, scientific computing, and real time analytics \cite{b29}. However, SSD behavior in the field varies with environmental conditions, workload characteristics, and design choices across vendors and product generations.

% \begin{figure}[!hbt]
%   \centering
%   % a) Smaller left figure
%   \begin{subfigure}[t]{0.15\textwidth}
%     \centering
%     \includegraphics[width=\linewidth]{figures/mot1.png}
%     \caption{}
%     \label{fig:failed_share}
%   \end{subfigure}
%   \hfill
%   % b) Larger right figure
%   \begin{subfigure}[t]{0.31\textwidth}
%     \centering
%     \includegraphics[width=\linewidth]{figures/mot2.png}
%     \caption{}
%     \label{fig:rasr}
%   \end{subfigure}
%   \caption{Production evidence motivates principled analysis: (a) quarterly variation in fleet annualized failure rate; (b) a conceptual heatmap where workload, environmental, and design factors influence latency, throughput, error rate, and endurance.}
%   \label{fig:fleet_stats}
% \end{figure}

Environmental conditions strongly shape performance and endurance as temperature, humidity, and vibration vary across sites and seasons \cite{b1,b2}. Airflow and cooling create rack level thermal and moisture gradients, neighboring servers introduce platform vibration, and weather can push facilities beyond their preferred operating envelope. These stressors raise latency and error rates and can trigger incidents that propagate across the fleet \cite{b3,b7,b11,b12,b13,b14,b21}. Controlled studies confirm that environmental stress elevates tail latency, reduces bandwidth, and increases the frequency of correctable and uncorrectable errors \cite{b1,b2,radiation}. Workload characteristics modulate drive behavior since request size, queue depth, randomness, compressibility, and locality impose pressure on buffering, logical to physical mapping, and garbage collection \cite{b20}. Hardware and architecture add further freedom because flash technology and bits per cell shape program and erase error modes \cite{b11,irps2024_tlc_qlc,retention_model_2024}, controller design influences scheduling and protection \cite{b5,b6,autoblox_micro23,ssd_iq_vldb2025}, and the interface controls command handling and parallelism \cite{b35,ibm_nvme_queues_2023,esf_nvme_2023}. Caching strategy and power loss protection can shelter applications while introducing new tradeoffs \cite{msst24_cache_coord,intelmem_plp_2023}, and firmware policies for error correction, garbage collection, wear leveling, and over provisioning couple these elements into complex dynamics \cite{b5,ren2025_ssd_survey,alahmadi2023_crash_recovery}.

These factors shape SSD performance and reliability as production views report quarterly drift in annualized failure rates, which rise from 0.96\% in 2023 Q1 to 1.05\% in 2023 Q2 \cite{backblaze_ssd_midyear_2023, backblaze_q1_2023, backblaze_q2_2023}, and studies \cite{b1,b2,b5,b6,b10,b20,b26,b31,b32} show that workload, environment, and design influence latency, throughput, error rate, and endurance.
% Figure~\ref{fig:fleet_stats} shows production views where the left panel reports quarterly drift in annualized failure rates and the right heatmap summarizes how workload, environment, and design influence latency, throughput, error rate, and endurance. 
This motivates continuous structured analysis at device and fleet scale so that telemetry becomes timely insights and decisions that sustain target performance and reliability. Continuous monitoring reveals early drift, supports diagnosis of root causes \cite{llm3,llm4,log_llm}, and guides interventions that stabilize latency, sustain throughput, and reduce error rates \cite{graf_ton2024,slopower_ccgrid24,arcus_2024}. Operators can rebalance workloads across racks, adjust thermal policy during rising temperature, schedule firmware updates during low demand, and replace at risk drives before customer impact. These tasks remain challenging because the factors are interdependent, interactions are often nonlinear, and key signals evolve over time, while SMART indicators offer limited exploratory insight \cite{b37}. Many operational questions are distributional. An engineer may need to predict whether the \(99^{\mathrm{th}}\) percentile latency next week will exceed a threshold during evening peak if rack temperatures rise by \(2^{\circ}\mathrm{C}\) \cite{graf_ton2024,slopower_ccgrid24}.

Prior work addresses parts of this landscape through several complementary strands, as experimental campaigns \cite{b1,b2} deliberately vary temperature, humidity, and vibration in controlled settings in order to measure how tail latency and bandwidth respond to environmental stress and to quantify the margins at which performance begins to degrade, while statistical workload studies \cite{b20,b23} analyze the distribution of request sizes, queue depths, read write ratios, randomness, and locality to uncover correlations with performance variability and to suggest operational knobs that practitioners can tune during incident response. A parallel body of investigations \cite{b9,b42,b24,b30,mvtrf,msfrd} mines SMART attributes and controller error logs to relate health indicators to impending failures and to shifts in performance, which enables threshold based alarms and risk scoring in production fleets, and operators regularly publish field reports \cite{b3,b14} that distill lessons from large telemetry repositories through careful manual inspection of health counters and workload traces in order to guide firmware upgrades, cooling policies, and replacement planning. Building on these descriptive efforts, many studies train machine learning models \cite{b28,b42,mvtrf} on SMART features to predict failures or to assign risk scores, and in stable lab datasets these models often achieve attractive accuracy, which has motivated repeated attempts to bring such predictors into production environments. Extending this line of work, machine learning predictors \cite{b28,b10} combine long term and short term features \cite{mvtrf}, and rating methods compare mutation patterns of live drives with historical cohorts \cite{msfrd}; cross layer studies argue that models should join SMART, controller counters, and NAND wear indicators because SMART alone can miss important device phenomena \cite{b11,b23,b24,b25}; and generative or unsupervised detectors such as VAE, GAN, and autoencoders flag anomalies but often return scores that provide little guidance to operators \cite{deeplog_2017,loggan_2021,vaelog_2023}.

Despite these advances, operational SSD telemetry analysis remains challenging. Manual workflows are time consuming, and machine learning pipelines often depend on ad-hoc feature selection. Training a joint model over heterogeneous telemetry is hard because labeled data are scarce: some signals are static, such as flash type and firmware algorithms, while others are dynamic, such as SMART, workloads, and environmental factors. As a result, model based approaches face heterogeneous and limited data, distribution shift across platforms and generations, and sparse labels for rare failures, which limits generalization. Many models cannot answer what-if questions or provide prescriptive recommendations that respect device constraints. Black box predictors also fail to show the facts behind their decisions. Experimental studies of environmental effects require specialized chambers and long campaigns, which makes them costly and slow to repeat, while proprietary traces and labels restrict open datasets, reproducibility, and progress. Finally, static fusion of SMART metrics, performance logs, and NAND errors struggles to separate workload noise from true degradation and offers little actionable guidance. There is a need for a framework that can analyze scattered telemetry holistically, deliver accurate predictive results, explain its reasoning with facts, produce clear descriptive summaries, offer prescriptive guidance to operators, and support what-if questions to aid decision-making. 

One promising solution is to use Large Language Models (LLMs) \cite{llm1,llm2,llm3,llm4,llm12,RAG}, which can read diverse documents, connect evidence, and produce explanations aligned with engineering reasoning \cite{llm2,graphrag_2024, ssd_llm_akewar}. They can integrate research papers, white papers, postmortems, and facility logs, translating operator questions into structured steps \cite{toolformer_2023}. However, applying LLMs to storage telemetry poses challenges: free-form generation can hallucinate \cite{hallucination, DBLP:conf/ipps/EgersdoerferSBB25}, correct reasoning requires rich context on device design, firmware, and operational constraints \cite{lost_middle_2023}, and long time series must preserve ordering, units, and align events across devices and sites. Spatial and temporal relations must capture interactions among workload, environment, and design rather than averaging them away. Scale is also a hurdle, as fleets comprise thousands of devices generating numerous measurement streams. Addressing these issues requires explicit grounding, careful retrieval of authoritative evidence, and representations that make reasoning auditable and repeatable.

Motivated by the potential of LLMs to analyze SSD operational data in an effective, interpretable, and exploratory manner, and aiming to overcome the limitations of prior approaches while enabling accurate, hallucination-resistant analysis, we define the following \textbf{Research Questions (RQs)}: \textit{ 
\textbf{RQ1:} How can we collect and structurally represent established facts and results from existing studies so that they are reusable during operation. 
\textbf{RQ2:} How can we represent operational telemetry so that a model can summarize behaviour across space and time for individual drives and for entire fleets. 
\textbf{RQ3:} How can we provide the right context to reduce hallucination and to improve the faithfulness of answers. 
\textbf{RQ4:} How can we deliver grounded and complete exploratory analysis that spans descriptive, predictive, prescriptive, and what-if tasks.}

\textbf{Our solution.} We address these questions with a knowledge orchestrated reasoning framework that combines LLM with Knowledge Graph (KG) \cite{kg1} grounded in an SSD taxonomy, in which literature and field evidence are distilled into entities, relations, and constraints aligned with a shared ontology \cite{kg1}, validated for consistency, and maintained as a living graph. Given a user query and recent operational data, the system issues a graph query to retrieve the pertinent subgraph and context, then generates evidence based explanations that support exploratory analysis for both device and fleet level operation, providing traceable reasoning and timely decision support even when data are sparse, heterogeneous, or partially missing. Our work makes the following contributions.

\textbullet \textit{A knowledge orchestrated reasoning framework that unifies descriptive, predictive, prescriptive, and what-if analysis for SSD operation through the combination of a language model and a knowledge graph grounded in an SSD vocabulary.}

\textbullet \textit{The first Literature KG focused on SSDs that is constructed from literature and operational sources and that is released to support reproducible research.}

\textbullet \textit{Methods for concept extraction, taxonomy alignment, and fact validation that deliver transparent answers with citations and work with limited, heterogeneous data.}

\textbullet \textit{An evaluation on production telemetry and diverse flash technologies that demonstrates expert level diagnosis and recommendations with markedly lower analysis time.}

% Evaluation uses production telemetry and field studies. We measure accuracy, usefulness, and explanation quality with metrics for correctness of descriptive summaries, precision of predictive alerts, impact of prescriptive recommendations on service quality, and fidelity of what if reasoning to known behavior. 
To our knowledge \textbf{KORAL} is the first to deliver full spectrum SSD reasoning by combining an LLM with a data KG and it includes the first literature KG dedicated to SSDs.
\section{Design Thinking}
\label{sec:bg}

Operators need a method that turns fragmented SSD telemetry into answers that are faithful to evidence and easy to act on. We address this need by combining an intermediate representation with a domain KG so that a LLM can reason over grounded facts rather than raw streams.

%\subsection{Motivation}
\begin{figure}[!hbt]
\centering
\includegraphics[width=\linewidth]{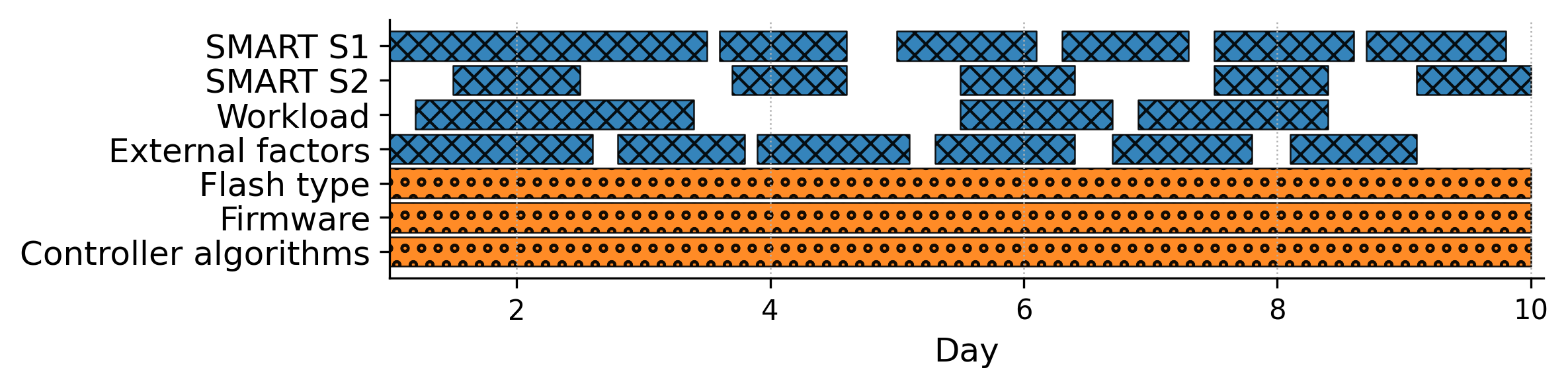}
% \caption{Timeline of SSD signals across days by data type, with dynamic signals changing daily, static descriptors staying constant, and both affecting behavior differently.}
\caption{Timeline of SSD signals across days by telemetry type. Cross-hatched bars denote dynamic, time-varying SMART telemetry (e.g., workload and external factors), while dotted bars denote static descriptors (e.g., flash type, firmware).}

\label{fig:ssd_timeline}
\end{figure}
\vspace{-0.2cm}
\subsection{Elements}
\subsubsection{Data}
SSD telemetry arrives in many forms and at different timelines. Homogeneous inputs include SMART attributes for a single drive over several days, which in our production dataset are sampled \emph{daily} (the sampling interval is adjustable), workload features such as queue depth, request size, and burstiness, and environmental factors like \(60^{\circ}\mathrm{C}\) temperature and \(50\%\) humidity \cite{dataset_details,b20,b2,b26}. Heterogeneous inputs combine sources, for example SMART with workload for a drive on a given day, SMART with environmental factors for a rack over two weeks, or SMART with workload and garbage collection traces for a busy application tier \cite{b23,b24,b5}.

Signals appear at multiple granularities such as per drive, per drive per day, per drive across \(n\) days, fleet level per day, and fleet level across \(n\) days (Figure~\ref{fig:ssd_timeline}). Fields may be dynamic such as reallocated sector count, program fail rate, write amplification, and duty cycle, or static such as flash type, controller model, interface type, and firmware version. This diversity introduces mixed granularity, temporal misalignment, schema drift, and missing values that can obscure true relationships. A one day temperature spike cannot be compared to a fourteen day fleet average, and raw streams given to a LLM can yield ambiguous prompts where missing values are treated as zeros.

We address these issues by transforming telemetry into a stable intermediate representation of summaries that record mean, \(95^{\text{th}}\) percentile, trend slopes, change points, correlations, uncertainty intervals, and explicit coverage or missing data flags. For missing values, we apply simple imputation for common attributes with short gaps, attach explicit DataQuality nodes in the knowledge graph that mark missingness and imputation decisions, and downweight or drop windows with severe gaps so downstream reasoning does not treat missing as observed. We add semantic links among drives, sites, firmware, and environmental context to form a structured knowledge graph that preserves context, normalizes heterogeneous inputs, and enables scalable queries, and language models then reason over these compact grounded facts rather than raw logs.

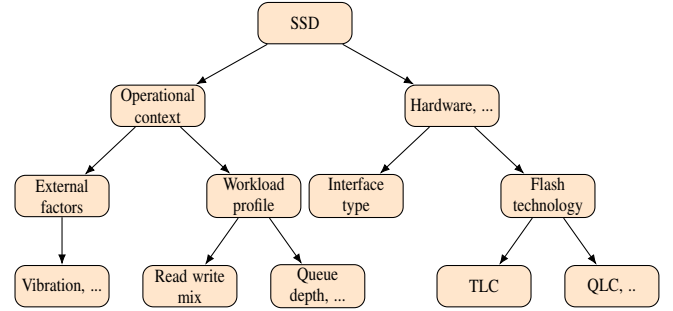
\begin{figure}[!hbt]
  \centering
  \begin{tikzpicture}[
      xscale=0.85,
      every node/.style={transform shape},
      taxnode/.style={draw, rounded corners, fill=orange!20,
                      inner sep=1pt, font=\scriptsize, align=center,
                      text width=14mm, minimum height=5.5mm},
      level 1/.style={sibling distance=46mm, level distance=11mm},
      level 2/.style={sibling distance=30mm, level distance=12mm},
      level 3/.style={sibling distance=20mm, level distance=12mm},
      edge from parent/.style={draw, -{Latex[length=1.4mm]}}
    ]
    \node[taxnode] {SSD}
      child { node[taxnode] {Operational\\context}
        child { node[taxnode] {External\\factors}
          child { node[taxnode] {Vibration, ...} }
        }
        child { node[taxnode] {Workload\\profile}
          child { node[taxnode] {Read write\\mix} }
          child { node[taxnode] {Queue\\depth, ...} }
        }
      }
      child { node[taxnode] {Hardware, ...}
        child { node[taxnode] {Interface\\type} }
        child { node[taxnode] {Flash\\technology}
          child { node[taxnode] {TLC} }
          child { node[taxnode] {QLC, ..} }
          % child { node[taxnode] {SLC} }
        }
      }
      % child { node[taxnode] {Firmware}
      %   child { node[taxnode] {Wear\\leveling, \\ ...} }
      %   child { node[taxnode] {Garbage\\collection} }
      % }
      % child { node[taxnode] {Metrics}
      %   child { node[taxnode] {Latency\\p99} }
      %   child { node[taxnode] {Throughput,\\...} }
      % };
      ;
  \end{tikzpicture}
  \caption{A slice of the SSD taxonomy used as shared vocabulary. The taxonomy is seeded by experts; KORAL may propose new concepts from literature, which are added after validation.}
  \label{fig:tax_tree}
\end{figure}

\subsubsection{Natural language and LLMs}
LLMs excel at open vocabulary queries, pattern abstraction, hypothesis generation, and clear explanations \cite{RAG,llm2, ssd_llm_akewar}. They can integrate logs and metrics, learn patterns without task specific code, and support natural language queries for operators. Prior work demonstrates effective use for log analysis \cite{b40,log_llm}, anomaly detection \cite{log_llm}, root cause analysis \cite{llm3,llm4}, scheduling and resource management \cite{llm6}, and storage tuning including hybrid SSD configuration \cite{b33,b34,llm_ssd}. Prompt strategies and in context learning such as LogPrompt \cite{b40} and AdaParser \cite{adaparser_2024} improve parsing and diagnosis, hybrid designs enrich events and fine tune smaller language models such as BERT \cite{sentenceEmbeddings} as in LogLLM \cite{log_llm}, and agent based studies explore root cause analysis with tool augmented LLMs \cite{llm3,llm4}. Knowledge graph reasoning and retrieval augmented generation connect signals, assets, and documents and recent surveys describe how these methods support global questions and explainable answers \cite{RAG}, while recent Graph RAG approaches explicitly build entity graphs and community summaries to answer global questions over large corpora \cite{kgroot_2024,graphrag_2025}. 

Direct application to raw telemetry remains challenging because outputs depend on prompt design and evidence order, advanced prompting strategies \cite{b41,tot_yao2023,got_besta2023} require careful tuning, and meta prompt optimization increases cost, latency, and variability. Risks such as hallucination, context limits, privacy concerns, and data drift \cite{apo_2023,lost_middle_2024,hallu_survey_2023} further complicate production use. By reasoning over a structured knowledge graph, LLMs can produce faithful and hallucination resistant analysis from simple prompts, which enables descriptive, predictive, prescriptive, and what-if queries efficiently and reliably.
% LLMs excel at open-vocabulary queries, pattern abstraction, hypothesis generation, and clear explanations \cite{RAG,llm2}. They can integrate logs and metrics, learn patterns without task-specific code, and support natural language queries for operators. Prior work demonstrates effective use for log analysis \cite{b40,log_llm}, anomaly detection \cite{log_llm}, root cause analysis \cite{llm3,llm4}, scheduling and resource management \cite{llm6}, and storage tuning including hybrid SSD configuration \cite{b33,b34,llm_ssd}.  
% %
% Direct application to raw telemetry remains challenging: outputs depend on prompt design and evidence order, advanced prompting strategies \cite{b41,tot_yao2023,got_besta2023} require careful tuning, and meta prompt optimization increases cost, latency, and variability. Risks such as hallucination, context limits, privacy concerns, and data drift \cite{apo_2023,lost_middle_2024,hallu_survey_2023} further complicate production use. By reasoning over the structured KG, LLMs can produce faithful, hallucination-resistant analysis from simple prompts, enabling descriptive, predictive, prescriptive, and what-if queries efficiently and reliably.

\subsection{Context}
\subsubsection{SSD taxonomy}
In this work we use a domain taxonomy of SSD together with a KG in order to support structured reasoning. The taxonomy supplies a shared vocabulary for devices, environments, workloads, and health signals, and the graph turns these concepts into entities and relations that can be queried, validated, and summarized in natural language. The taxonomy organizes knowledge that operators already use in practice. At the top level it includes the operational context with external factors such as temperature, humidity, and vibration, the workload profile with read write mix and queue depth, the hardware stack with flash technology and interface type, the firmware layer with error correction, wear leveling, and garbage collection, and the metrics that describe performance and reliability. Figure~\ref{fig:tax_tree} sketches a portion of the taxonomy that we use as a running vocabulary in the framework. The taxonomy is seeded by SSD domain experts, and KORAL only extends it by proposing out of vocabulary concepts automatically and adding them to the taxonomy after manual validation.

\subsubsection{Ontology and knowledge graph}
An ontology gives formal meaning to the vocabulary and constrains how concepts can be combined. We model an ontology as
\[
\mathcal{O} = \big(C,\; R,\; I,\; A\big),
\]
where \(C\) is a set of classes, \(R\) is a set of binary relations with declared domain and range, \(I\) is a set of individuals, and \(A\) is a set of axioms that include subclass assertions, typing assertions, and integrity constraints. In practice we express \(\mathcal{O}\) with classes such as \(\textit{EnvironmentalFactor}\), \(\textit{WorkloadAttribute}\), \(\textit{OperationType}\), and \(\textit{Metric}\), and with relations such as \(\textit{impactsOperation}\), \(\textit{impactsMetric}\), \(\textit{greaterImpactThan}\), and \(\textit{hasContext}\).
A KG is a set of typed entities connected by labeled edges that satisfy the ontology. Formally we write
\[
\mathcal{G} = \big(V,\; E,\; \tau_V,\; \tau_E\big),
\]
where \(V\) is a set of nodes, \(E \subseteq V \times R \times V\) is a set of labeled edges, \(\tau_V: V \rightarrow C\) assigns a class to each node, and \(\tau_E: E \rightarrow R\) assigns a relation to each edge. When stored in Resource Description Framework form the graph is a set of triples \(T \subseteq U \times U \times U\) over internationalized resource identifiers and literals, with semantics governed by RDFS \cite{rdfs11} and OWL \cite{owl2-overview}.

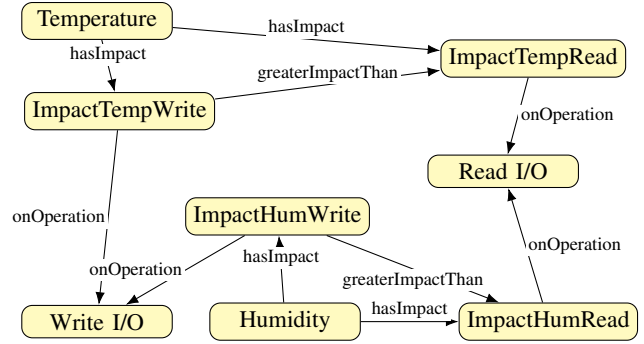
\begin{figure}[!hbt]
  \centering
  \footnotesize
  \begin{tikzpicture}[
      >=Latex,
      entity/.style={draw, rounded corners, fill=yellow!30, inner sep=3pt,
                     minimum width=20mm, minimum height=4mm, font=\small, align=center},
      lab/.style={font=\scriptsize, fill=white, inner sep=1pt}
    ]
    \node[entity] (temp) at (0,3)  {Temperature};
    \node[entity] (hum)  at (2.5,-1)  {Humidity};
    \node[entity] (write) at (0,-1) {Write I/O};
    \node[entity] (read)  at (5.4,1) {Read I/O};
    \node[entity] (itw) at (0.3,1.8) {ImpactTempWrite};
    \node[entity] (itr) at (5.8,2.5) {ImpactTempRead};
    \node[entity] (ihw) at (2.4,0.4) {ImpactHumWrite};
    \node[entity] (ihr) at (6,-1) {ImpactHumRead};
    \draw[->] (temp) -- (itw) node[midway,lab,above] {hasImpact};
    \draw[->] (temp) -- (itr) node[midway,lab,above] {hasImpact};
    \draw[->] (itw)  -- (write) node[midway,lab,left]  {onOperation};
    \draw[->] (itr)  -- (read)  node[midway,lab,right] {onOperation};
    \draw[->] (itw)  -- (itr)   node[midway,lab,above] {greaterImpactThan};
    \draw[->] (hum) -- (ihw) node[midway,lab,above] {hasImpact};
    \draw[->] (hum) -- (ihr) node[midway,lab,above] {hasImpact};
    \draw[->] (ihw) -- (write) node[midway,lab,left]  {onOperation};
    \draw[->] (ihr) -- (read)  node[midway,lab,right] {onOperation};
    \draw[->] (ihw) -- (ihr)   node[midway,lab,below] {greaterImpactThan};
  \end{tikzpicture}
  \caption{Automatically generated Literature KG fragment from Stage I that encodes the claim that \textit{"temperature and humidity impact write I/O more than read I/O"}. Nodes and relations are produced by the automated extraction and ontology alignment pipeline, with provenance linked to the supporting sentence.}

  \label{ontology_example}
\end{figure} 
\textbf{Example graph.}
Consider the statement that \textit{temperature and humidity changes impact the write I/O operations more than the read I/O operations}. Figure~\ref{ontology_example} shows one encoding that uses the taxonomy terms for temperature, humidity, and operations and the ontology relations introduced above. The nodes \(\textit{ImpactTempWrite}\) and \(\textit{ImpactTempRead}\) represent specific impact instances and the edge \(\textit{greaterImpactThan}\) captures the comparative claim.

% \subsubsection{Storing and querying the graph}
% We store the graph in Turtle to support version tracking and auditing and to ensure portability across tools using standard RDF and RDFS vocabularies \cite{rdfs11}. We query the graph with SPARQL \cite{rdfs11, owl2-overview} to retrieve evidence for analysis; due to page limits, we omit the query and graph storing structure snippets.

\subsubsection{Storing and querying the graph}
We store the graph in the Turtle format so that versions can be tracked and audited and so that the graph remains portable across tools. The snippet below mirrors Figure~\ref{ontology_example} and uses standard RDF and RDFS vocabulary \cite{rdfs11}.

{\footnotesize
\begin{verbatim}
@prefix ex: <http://example.org/ssd#> .
ex:Temperature ex:hasImpact ex:ImpactTempWrite .
ex:ImpactTempWrite ex:greaterImpactThan ex:ImpactTempRead.
\end{verbatim}
}

We query the graph with SPARQL \cite{rdfs11, owl2-overview} in order to retrieve evidence for analysis. The example below asks for environmental factors that have a stronger impact on write operations than on read operations and returns both impact nodes.

{\footnotesize
\begin{verbatim}
PREFIX ex: <http://example.org/ssd#>
SELECT ?factor ?impactWrite ?impactRead
WHERE {?factor a ex:EnvironmentalFactor .
       ?factor ex:hasImpact ?impactWrite .
       ?factor ex:hasImpact ?impactRead .
       ?impactWrite ex:onOperation ex:WriteIO .
       ?impactRead  ex:onOperation ex:ReadIO .
       ?impactWrite ex:greaterImpactThan ?impactRead.}
\end{verbatim}
}

The same pattern can be used with performance metrics. For example the relation \textit{impactsMetric} can connect temperature to latency p ninety nine under a given workload profile, and a query can return the most relevant mitigation actions that are attached to the same subgraph.

\paragraph{Data quality validation.}
Currently, missingness and imputation decisions are represented explicitly as DataQuality nodes and are used to downweight or drop low coverage windows during analysis. As future work, we plan to encode these validation rules as SHACL constraints, enabling declarative conformance checking of Data KG nodes before they are consumed by summarization and prompting.

\section{KORAL Architecture}
\label{sec:method}

\begin{figure*}[!hbt]
\centering
\includegraphics[width=\linewidth
]{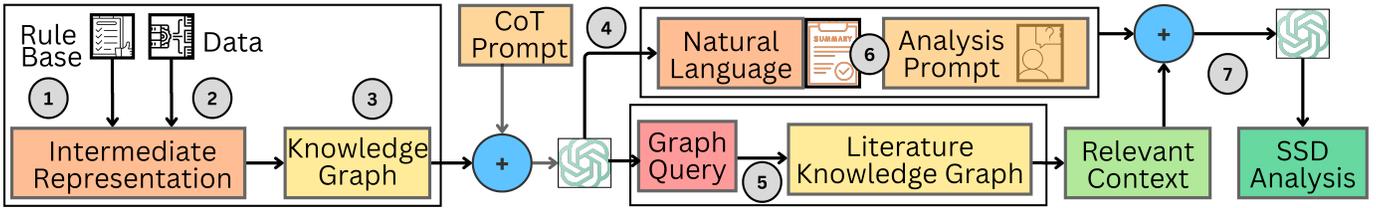}
\caption{KORAL Stage II Flow (RQ2, RQ3 \& RQ4). The only expert authored component is the curated rule repository that defines how telemetry is summarized into frames and mapped to graph relations. Given this rule base, intermediate representation construction, Data KG creation, node linking, summarization, retrieval, and generation are automated.}

\label{fig:stageII}
\end{figure*}

KORAL follows a two stage workflow (Figures~\ref{fig:stageI} and~\ref{fig:stageII}). Stage I builds a literature grounded knowledge graph aligned to an SSD taxonomy and ontology, with validation and controlled updates. Stage II transforms heterogeneous telemetry into a rule based intermediate representation, materializes a Data KG, retrieves relevant literature evidence, and generates grounded SSD analysis that supports descriptive, predictive, prescriptive, and what-if queries.

\subsection{Stage I: Building the Literature Knowledge Graph}
\label{subsec:stageI}

\begin{figure}[!hbt]
\centering
\includegraphics[width=\linewidth
]{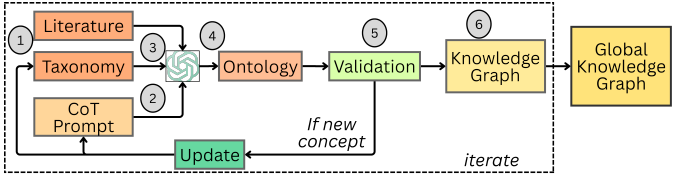}
\caption{KORAL Stage I Flow (RQ1). Document parsing, claim extraction, entity linking, and Literature KG construction are automated. When an out of vocabulary concept is detected, the system generates a concept proposal that is manually reviewed and validated before it is added to the vocabulary.}
\label{fig:stageI}
\end{figure}

Stage I constructs a Literature KG by extracting evidence backed claims from papers and reports and aligning them to a shared SSD vocabulary from the taxonomy and to typed relations from the ontology (Figure~\ref{fig:stageI}). The output is an append only graph where each assertion retains provenance, including the evidence sentence and the source document identifier, so later analysis can cite supporting text and surface disagreements rather than overwriting them.

\paragraph{Inputs and preparation}
We build a literature corpus from research and industry sources and convert each document into clean text with section anchors, so every extracted fact points to its supporting sentence. We pair this corpus with an SSD taxonomy that provides canonical anchors for external factors such as temperature and humidity, workload dimensions such as access pattern and queue depth, and metrics such as IOPS and the \(99^{\text{th}}\) percentile latency. Each document is assigned a stable identifier to ensure traceability during graph merging into a global KG.

\paragraph{Prompt guided extraction and ontology alignment}
A language model reads the full document under taxonomy guidance and returns a strict JSON containing entities, triples, mappings, axioms, and optional concept proposals. The prompt enforces taxonomy first mapping by linking each mention to the closest taxonomy path and preferring specific children such as \(99^{\text{th}}\) percentile latency over generic parents. It distinguishes classes from instances, requires explicit context that connects an SSD instance to environment and optionally to a workload profile, and uses directional predicates when a paper reports improvement or degradation. Each triple carries an evidence sentence and a confidence score to preserve provenance and to support auditing.

We align extracted concepts to ontology classes (for example EnvironmentalFactor, WorkloadAttribute, OperationType, and Metric) and relations (for example impactsOperation, impactsMetric, causes, mitigatedBy, and greaterImpactThan), and we normalize synonyms such as tail latency and p99 latency to the same metric class. This alignment ensures that queries over the graph remain consistent across sources and devices.

\paragraph{Validation, evidence alignment, and versioned materialization}
Stage I enforces a validation gate before any extracted claim is committed to the Literature KG. We validate each JSON output for schema completeness, apply ontology type checks to ensure every triple has compatible subject and object types, and align each triple to an explicit evidence span in the source text. Triples that fail schema or type constraints, lack evidence alignment, or fall below a configurable confidence threshold are rejected. Units and percentiles are standardized where applicable, and contradictory claims are retained with distinct provenance but down weighted during retrieval so higher confidence, better supported evidence is preferred. Validated triples are written in Turtle form and versioned in a repository, and we maintain both per collection graphs and a merged global graph so knowledge accumulates safely over time.

\paragraph{Worked example}
Consider the sentence that a high inlet temperature raises the \(99^{\text{th}}\) percentile latency under a write heavy workload. The extractor maps this to the entities Temperature, \(99^{\text{th}}\) percentile latency, and Write heavy, and encodes a relation that connects the environmental factor to the latency metric under the specified workload condition. Under the prompt constraints, the claim becomes a typed assertion with a directional predicate (for example \texttt{degrades}), includes the evidence sentence verbatim, and binds the entities to taxonomy paths so that later graph queries can retrieve and reuse the claim consistently.

\paragraph{Vocabulary growth under controlled review}
When a document introduces a concept not covered by the current taxonomy, the prompt asks the model to propose a new concept with a short definition and a suggested parent. The pipeline records these proposals for review, so the vocabulary can evolve without breaking existing mappings and without silently changing meaning across versions.

\paragraph{Outcome}
Stage I yields a Literature KG that distills entities, relations, constraints, and provenance from the corpus, aligned to the SSD taxonomy and validated for consistency. This graph serves as the reusable evidence base for Stage II during operational analysis.

\subsection{Stage II: Operational SSD Analysis}
\label{subsec:stageII}

Stage II begins from the available telemetry and selects an intermediate representation using a rule base (Figure~\ref{fig:stageII}). The rule base is an expert curated repository of summarization and mapping rules, and it is not automatically learned from data. The rule base defines how to summarize dynamic signals such as SMART, workload, and environmental factors across drive and fleet granularity within an \(n\) day window, while passing static descriptors such as flash technology, controller model, and firmware as stable facts. The goal is to preserve temporal structure and data quality information while producing compact, typed frames that can be turned into a Data KG.

\paragraph{Intermediate representation and frame emission}
The pipeline computes robust statistics and temporal features such as trends, rate of change, change points, seasonality, correlation, exposure above thresholds, and explicit coverage and missingness indicators. It emits typed frames with stable identifiers and provenance, including AttributeFrame for SMART summaries, WorkloadFrame for workload profiles, EnvFrame for environmental summaries, and DataQuality frames that record missingness, sensor checks, and imputation. It also emits Episode or Observation frames when multi day regime shifts or cross source associations are detected.

\paragraph{Worked example}
For a drive window, the system can create an AttributeFrame for percentage used that includes median, \(95^{\text{th}}\) percentile, rate of change per day, Mann Kendall trend statistic and p value. It can also create a WorkloadFrame with read share, average queue depth, and burstiness, and a DataQuality frame that reports partial coverage such as five observations out of seven with one imputed value. If a change point detector flags a regime shift starting on day four, the pipeline records this as an Episode linked to the same window identifier so downstream reasoning can separate transient spikes from persistent drift.

\paragraph{Materializing the data knowledge graph}
We build the Data KG by converting frames into typed nodes with window tags, computation methods, confidence, and provenance, and by instantiating relation templates from the rule base. For example, SMART summaries become AttributeFrame nodes connected to the corresponding drive and window, workload and environment frames connect through hasContext style edges, and Episodes link to the signals that triggered them. This graph provides a compact operational context that replaces raw time series during downstream analysis while keeping traceability intact.

\paragraph{Grounded summarization, retrieval, and answer generation}
Given the Data KG, KORAL generates a natural language summary that preserves units, time windows, uncertainty, and data quality statements. In parallel, it produces a graph query over the shared vocabulary to retrieve a relevant subgraph from the Literature KG. The final prompt combines the user question, the Data KG grounded summary, and the retrieved literature context, then generates descriptive, predictive, prescriptive, or what-if outputs depending on the query intent and available frames.

\paragraph{Audit trail}
To support reproducibility and debugging, KORAL stores the emitted frames, the Data KG identifiers, the literature query, retrieved evidence, and the final response as artifacts tied to the same window tag, enabling later inspection of what was observed, what was retrieved, and what the model claimed. When the operator accepts a recommendation the system can attach a mitigation action node to the data graph for that window and can schedule a follow up query that checks whether the expected improvement in latency and error rates occurred during the next week, which closes the loop between observation, literature grounded reasoning, and operational action.

\paragraph{Automation versus expert authored components}
KORAL is fully automatic at runtime. The only expert authored artifact is the curated rule repository, which specifies attribute groupings, ideal directions, windowing and summarization functions, and the mapping from frames to typed graph relations. All other modules, including intermediate representation construction, Data KG materialization, node linking, summarization, literature retrieval, and response generation, execute automatically given telemetry and the rule repository. In Stage I, the pipeline can automatically flag out of vocabulary concepts, but they are only added to the vocabulary after manual validation.

\subsection{Running Example: From Telemetry to Prompt}
\label{subsec:stageII_running_example}
Figure~\ref{fig:stageII} outlines Stage II at a high level. We now trace a single drive through the full pipeline using a 30 day SMART window (daily samples) and a representative subset of attributes. Each SMART attribute is also tagged with an \emph{ideal} direction (Low, High, or Monitor) to preserve polarity for later scoring and explanations.

\paragraph{Raw telemetry (30 day SMART window)}
For a drive \(d\), we ingest a daily SMART record \(x_t\) containing counters such as reallocated sectors (\texttt{r\_5}), wear and erase indicators (\texttt{r\_173}, \texttt{r\_177}), program and erase failures (\texttt{r\_181}, \texttt{r\_182}), uncorrectable errors (\texttt{r\_187}), pending sectors (\texttt{r\_197}), CRC errors (\texttt{r\_199}), and traffic counters such as total LBAs written and read (\texttt{r\_241}, \texttt{r\_242}). The rule base also records coverage (how many days observed), missingness, and simple validity checks (for example counter monotonicity).

\paragraph{Intermediate representation (SMART attribute frames)}
The rule base converts each attribute time series \(\{x_t\}_{t=1}^{30}\) into an \emph{AttributeFrame} that preserves temporal structure and uncertainty in a compact form. Each frame stores robust summaries and trend descriptors, for example median, tail percentiles, slope, volatility, change points, and exposure beyond operator meaningful thresholds, together with metadata such as the attribute id, units, and the ideal direction. A schematic example frame is shown below.

{\footnotesize
\begin{verbatim}
AttributeFrame(
  drive_id = "Disk/26871",
  window   = "2019-11-09..2019-12-08",
  attr_id  = "r_187", % Reported_Uncorrectable_Errors
  ideal    = "Low",
  summary  = {median, p95, last, max},
  temporal = {slope, change_point?, spike_days?},
  quality  = {coverage=30/30, missing_days=0})
\end{verbatim}
}

\begin{figure}[!t]
\centering
\footnotesize
\begin{tikzpicture}[
  node distance=4mm and 5mm,
  box/.style={draw, rounded corners, align=center, inner sep=2pt},
  edge/.style={-Latex, line width=0.4pt},
  lab/.style={font=\scriptsize}
]
\node[box] (drive) {Drive\\\texttt{Disk/26871}};
\node[box, right=15mm of drive] (win) {Window\\\texttt{2019-11-09..2019-12-08}};
\draw[edge] (drive) -- node[lab, above]{\texttt{hasWindow}} (win);

\node[box, below left=5mm and 10mm of win] (a187) {AttrFrame\\\texttt{r\_187}\\(uncorr.\ errors)};
\node[box, below=9mm of win] (a5)   {AttrFrame\\\texttt{r\_5}\\(realloc.\ sectors)};
\node[box, below right=6mm and 1mm of win] (a241) {AttrFrame\\\texttt{r\_241}\\(LBA \\ written)};
\draw[edge] (win) -- node[lab, left]{\texttt{hasAttrFrame}} (a187);
\draw[edge] (win) -- node[lab, left]{\texttt{hasAttrFrame}} (a5);
\draw[edge] (win) -- node[lab, right]{\texttt{hasAttrFrame}} (a241);

\node[box, below left=10mm and 45mm of a241] (dq187) {DataQuality\\coverage, missingness};

\draw[edge] (a187) -- node[lab, above]{\texttt{hasQuality}} (dq187);

\node[box, below left=10mm and 1mm of a241] (gmed) {Group\\\texttt{media}};
\node[box, below=15mm of a241] (guse) {Group\\\texttt{usage}};
\draw[edge] (a5) -- node[lab, above]{\texttt{belongsToGroup}} (gmed);
\draw[edge] (a241) -- node[lab, above]{\texttt{belongsToGroup}} (guse);

\node[box, below left=15mm and -18mm of a5] (ideal) {IdealDirection\\Low or High or Monitor};
\draw[edge] (a5) -- node[lab, left]{\texttt{hasIdeal}} (ideal);
\end{tikzpicture}
\vspace{-1mm}
\caption{Schematic Data KG for one drive and a 30 day SMART window. Applying the curated rule repository to raw telemetry automatically creates AttributeFrame and DataQuality nodes and their links, enabling traceable claims with units, window scope, and coverage.}
\label{fig:dataKG_schematic}
\vspace{-2mm}
\end{figure}
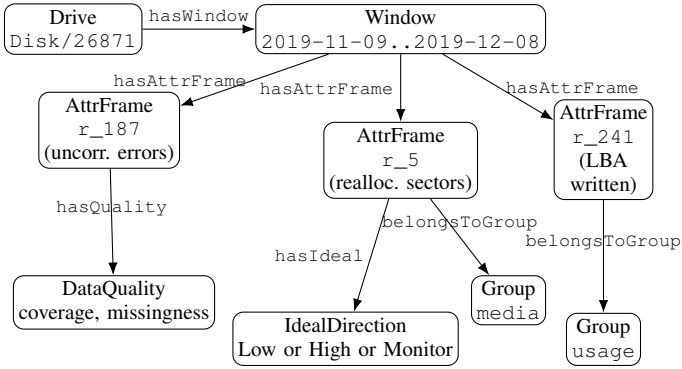

\paragraph{Data KG materialization (how the graph looks)}
Frames are then materialized into a Data KG where each node is typed and each edge is auditable. Figure~\ref{fig:dataKG_schematic} shows a readable subgraph for one drive and one window. The drive node links to the window node; the window links to AttributeFrames for selected SMART attributes; each attribute frame links to a data quality node, and to a semantic group node (media, interface, usage) so later retrieval and prompting can reference families rather than long attribute lists. This structure is what enables traceability, because every generated claim can be mapped back to a specific frame and window.

\paragraph{LLM prompt assembly (what the model actually sees)}
Finally, KORAL converts the Data KG subgraph into a compact, structured context and optionally augments it with retrieved evidence from the Literature KG using the same vocabulary (attribute names, workload concepts, and metrics). The prompt includes (i) the user query, (ii) the Data KG summary with units and coverage statements, and (iii) retrieved literature snippets with citations. A simplified prompt skeleton is shown below.

{\footnotesize
\begin{verbatim}
Task: Provide a grounded SSD analysis for Disk/26871 
over window 2019-11-09..2019-12-08.
Data KG context (auditable):
- r_187 Reported_Uncorrectable_Errors: trend=?, 
change_point=?, coverage=30/30
- r_5   Reallocated_Sector_Count: trend=?, 
spikes=?,  coverage=30/30
- r_241 Total_LBA_Written:  
burstiness=?, coverage=30/30
- Data quality: missing_days=0, notes=...
Retrieved literature (if available):
- Evidence linking thermal or workload
stressors to tail latency or error behavior, 
with citations.
User query:
"Explain the risk signals and suggest mitigations. 
Provide a what if comparison if workload is reduced."
\end{verbatim}
}

This running example illustrates how Stage II transforms a 30 day stream of SMART counters into a structured intermediate representation, materializes it as a typed Data KG with window level provenance, and then converts the relevant subgraph into a compact prompt context that can be augmented with retrieved literature evidence for grounded analysis.

\section{Evaluation}
\label{sec:evaluation}
This section defines the datasets, evaluation protocol, metrics, and results. We evaluate \textsc{KORAL} on individual SSD telemetry data, which includes SMART attributes \cite{b24, b29}, workload counters\cite{b29, b36}, and environmental measurements \cite{b1,b2}. We also assess heterogeneous combinations of these signals by augmenting dynamic telemetry with static device metadata. We evaluate performance and reliability at device and fleet levels in an exploratory analysis covering descriptive, predictive, prescriptive, and what-if scenarios.

\subsection{Datasets and Setup}
\label{subsec:datasets_setup}

\textbf{SMART telemetry, workload context, and failures.}
We use two complementary SSD corpora. The Google field study spans millions of drive days across six years, about ten device models, and multiple flash types \cite{b24}, and we use it to guide attribute selection, quality checks, and long horizon drift handling \cite{b29}. The Alibaba dataset provides daily snapshots from January 2018 to December 2019 for more than five hundred thousand drives across six SSD models (MA1, MA2, MB1, MB2, MC1, MC2), with confirmed failure tickets, rack or site identifiers, and application tags \cite{b29}. Each snapshot reports up to 105 SMART attributes per drive, including vendor normalized scores \texttt{n\_i} and raw counters \texttt{r\_i}, together with device id, timestamp, and model code, enabling analysis of absolute error counts, wear indicators, and cross model differences \cite{b37,b29}. We keep the 15 raw attributes and treat each drive's history as a multivariate daily time series. We focus on the \textbf{\texttt{MB1}} and \textbf{\texttt{MB2}} models in our experiments because they have very few missing values for each of the 15 attributes and are also used by prior work for evaluation making it straightforward to compare.

\textbf{Environmental signals and modality alignment.}
We incorporate environmental time series from two controlled studies \cite{b1,b2} that report temperature in \(^{\circ}\mathrm{C}\), relative humidity in \(\%\), and vibration frequency and amplitude, together with measured throughput and tail latency percentiles. We preserve units and provenance fields (study, run, and configuration) and retain metadata such as device model, firmware revision, sampling interval, and exposure duration for reproducibility. We evaluate five modality settings: SMART plus Workload, SMART plus Environmental, Workload plus Environmental, SMART plus Workload plus Environmental, and the same joint set augmented with static descriptors (for example flash type and firmware). For SMART plus Workload we use Alibaba traces \cite{b29} whose workload taxonomy includes eight categories: Web Service Management (WSM), Resource Management (RM), Web Proxy Services (WPS), SQL Services (SS), Database (DB), Web Services (WS), Data Analytics Engine (DAE), and Network Attached Storage (NAS). Because few traces contain all signals for the same device and window, device level joint analysis uses paired or tripled windows by aligning Alibaba SMART and workload windows with the environmental conditions in \cite{b1,b2} using window keys and strict timestamp tolerance, then filtering by coverage and monotonicity checks. For Workload plus Environmental tests, we generate workload profiles using FIO \cite{b36} following published request size and concurrency settings and combine them with the environmental conditions from \cite{b1,b2}. For fleet level analysis we avoid synthetic joins and rely on natively aligned Alibaba SMART and workload telemetry \cite{b29}, and we do not incorporate environment at fleet scale because these feeds cannot be synchronized to the production traces.

\textbf{Protocol, prompts, and reference text.}
We follow the Stage II pipeline (Sec.~\ref{subsec:stageII}) to build the intermediate representation and Data KG, retrieve context from the Literature KG, and generate outputs using GPT 4o \cite{b39}. Reliability experiments use 30 day SMART windows with ground truth labels for failure within a fixed horizon and time to failure defined as days from the window end to the failure event. At the drive level, for each window we issue four prompts covering predictive, descriptive, prescriptive, and what-if tasks, then aggregate metrics across windows as defined in Sec.~\ref{subsec:eval-metrics}. At the fleet level, we ingest each trace as a cohort, treat predicted failing drives as individual predictions for classification against ground truth, and generate one predictive, one descriptive, and one what-if output per query, then aggregate metrics across query types. For natural language evaluation, we create reference summaries and rationales using GPT 4o \cite{b39} as a drafting assistant by supplying evidence passages from \cite{b29,b1,b2,b24} together with representative intermediate representations; we manually verify and edit all references for fidelity and remove unsupported statements.

\subsection{Evaluation Metrics}
\label{subsec:eval-metrics}
We evaluate KORAL along two axes: numeric prediction for reliability and performance, and LLM generated text for descriptive summaries, prescriptive recommendations, and what-if analysis. Results in Table~\ref{tab:eval-table1} and Table~\ref{tab:eval-table2} report classification and regression metrics for prediction, and BLEU and ROUGE plus grounding metrics for text.

\paragraph{Reliability prediction (classification)}
For each drive \(d\) and analysis window ending at time \(t\), the binary ground truth failure label \(Y^{\mathrm{fail}}_{d,t}\in\{0,1\}\) indicates whether the drive fails within the evaluation horizon after the window, and the model outputs \(\hat{Y}^{\mathrm{fail}}_{d,t}\in\{0,1\}\). We report \emph{Precision} \(\mathbf{P}=\frac{\mathrm{TP}}{\mathrm{TP}+\mathrm{FP}}\), \emph{Recall} \(\mathbf{R}=\frac{\mathrm{TP}}{\mathrm{TP}+\mathrm{FN}}\), and \emph{Accuracy} \(\mathbf{A}=\frac{\mathrm{TP}+\mathrm{TN}}{\mathrm{TP}+\mathrm{FP}+\mathrm{FN}+\mathrm{TN}}\) \cite{powers2011evaluation,sokolova2009systematic}. In our SSD setting, True Positive (TP) is a window correctly flagged as at risk for a drive that fails within the horizon, False Positive (FP) is an unnecessary alarm on a healthy drive window, False Negative (FN) is a missed warning for a drive that fails within the horizon, and True Negative (TN) is a correctly identified healthy window.

\paragraph{Reliability and performance estimation (regression)}
We also evaluate continuous targets with mean squared error. For time to failure, \(T^{\mathrm{ttf}}_{d,t}>0\) is the number of days from the end of window \(t\) until the recorded failure event, and we report \(\mathrm{MSE}_{\mathrm{TTF}}=\frac{1}{N}\sum_{i=1}^{N}\bigl(T^{\mathrm{ttf}}_{i}-\hat{T}^{\mathrm{ttf}}_{i}\bigr)^2\). For example, if a drive fails 18 days after the window ends but the model predicts 25 days, the squared error for that window is \((18-25)^2=49\) days\(^2\), and MSE averages this value across all evaluated windows. For performance, we predict tail latency \(L^{(p)}_{d,t}\) (the \(p\) percentile latency in milliseconds) and report \(\mathrm{MSE}_{\mathrm{TL}}=\frac{1}{N}\sum_{i=1}^{N}\bigl(L^{(p)}_{i}-\hat{L}^{(p)}_{i}\bigr)^2\). For example, if the observed \(p99\) latency under a given temperature and workload condition is 3.2 ms and the model predicts 2.6 ms, the squared error is \((3.2-2.6)^2=0.36\) ms\(^2\), and MSE averages this error across all windows and runs.

\paragraph{Text quality and grounding}
For descriptive, prescriptive, and what-if outputs, we measure surface overlap against curated references using BLEU 4 \cite{papineni2002bleu} and ROUGE L \cite{lin2004rouge}. BLEU 4 is computed from modified \(n\) gram precisions up to \(n=4\) with a brevity penalty, \(\mathbf{B4}=\mathrm{BP}\cdot\exp\!\bigl(\sum_{n=1}^{4} w_n\log p_n\bigr)\). ROUGE L is based on the longest common subsequence between a generated text and the reference and is reported as an \(F_1\) score \cite{lin2004rouge}. Because overlap does not guarantee factual correctness, we also report task aligned grounding metrics. For descriptive summaries and prescriptive recommendations we use Faithfulness Precision, \(\mathbf{FiP}=\frac{\#\text{supported atomic claims}}{\#\text{atomic claims}}\), where a claim such as ``Percentage\_Used rose by 12\% over 30 days'' counts only if it is verified from the intermediate representation or Data KG nodes for the same drive and window, including units and coverage. For what-if analysis we use Counterfactual Validity, \(\mathbf{CFV}=\frac{\#\text{counterfactual statements with evidence consistent direction}}{\#\text{counterfactual statements}}\), for example ``if inlet temperature decreases by \(5^{\circ}\mathrm{C}\), tail latency decreases'' counts only when cohort evidence or retrieved literature supports that directional association under comparable conditions.

\subsection{Results}
\newcommand{\NA}{\textemdash}
\begin{table*}[!hbt]
\centering
\caption{SSD Operational Data Analysis (\textbf{Drive Level}).
Abbreviations: \textbf{FT} = Flash Type, \textbf{Al} = Algorithms. \textbf{P} = Precision (\%), \textbf{R} = Recall (\%), \textbf{A} = Accuracy (\%), \textbf{TTF-MSE} = Time-to-Failure-Mean Square Error (days), \textbf{TL-MSE} = Tail Latency-Mean Squared Error (\%),
\textbf{B4} = BLEU-4, \textbf{RL} = ROUGE-L, \textbf{FiP} = Faithfulness Precision, \textbf{CFV} = Counterfactual Validity, \textbf{\_\_ = NA}.}
\label{tab:eval-table1}
\renewcommand{\arraystretch}{0.9}
\small
\setlength{\tabcolsep}{4.5pt}
\begin{tabularx}{\textwidth}{X X c c c c c c c c c c c c c c}
\toprule
\textbf{Dataset Type} & \textbf{Dataset Used} &
\multicolumn{5}{c}{\textbf{Predictive}} &
\multicolumn{3}{c}{\textbf{Descriptive}} &
\multicolumn{3}{c}{\textbf{Prescriptive}} &
\multicolumn{3}{c}{\textbf{What-if}} \\
\cmidrule(lr){3-7} \cmidrule(lr){8-10} \cmidrule(lr){11-13} \cmidrule(lr){14-16}
& & \makecell{\textbf{P}\\\textbf{$\uparrow$}} & \makecell{\textbf{R}\\\textbf{$\uparrow$}} & \makecell{\textbf{A}\\\textbf{$\uparrow$}} & \makecell{\textbf{TL-}\\\textbf{MSE$\downarrow$}} & \makecell{\textbf{TTF-}\\\textbf{MSE$\downarrow$}}
& \makecell{\textbf{B4}\\\textbf{$\uparrow$}} & \makecell{\textbf{RL}\\\textbf{$\uparrow$}} & \makecell{\textbf{FiP}\\\textbf{$\uparrow$}}
& \makecell{\textbf{B4}\\\textbf{$\uparrow$}} & \makecell{\textbf{RL}\\\textbf{$\uparrow$}} & \makecell{\textbf{FiP}\\\textbf{$\uparrow$}}
& \makecell{\textbf{B4}\\\textbf{$\uparrow$}} & \makecell{\textbf{RL}\\\textbf{$\uparrow$}} & \makecell{\textbf{CFV}\\\textbf{$\uparrow$}} \\
\midrule
SMART & Alibaba &
\textbf{96} & 41 & 92 & \NA & 20 &
65 & 73 & 79 &
62 & 71 & 74 &
59 & 70 & 67 \\
SMART & Google &
92 & 38 & 90 & \NA & 23 &
63 & 72 & 77 &
60 & 70 & 72 &
57 & 69 & 65 \\
Environmental (Env.) & Exp. Studies &
\NA & \NA & \NA & \textbf{2.9} & \NA &
64 & 73 & 78 &
62 & 71 & 74 &
66 & 73 & 79 \\
SMART+Workload & Alibaba &
94 & 49 & 93 & \NA & \textbf{15} &
68 & 75 & {82} &
{65} & {73} & {78} &
{63} & {72} & {75} \\
SMART+Env. & Alibaba + Exp. Studies &
{95} & {52} & \textbf{94} & 3.2 & 20 &
\textbf{69} & \textbf{76} & \textbf{83} &
{66} & \textbf{74} & \textbf{80} &
{66} & \textbf{74} & {79} \\
Workload+Env. & FIO + Exp. Studies &
\NA & \NA & \NA & 4.2 & \NA &
{65} & {74} & {79} &
{63} & {72} & {77} &
{62} & {71} & {76} \\
\textbf{SMART+Workload+Env.} & Alibaba + Exp. Studies &
\textbf{96} & \textbf{58} & \textbf{94} & 3.0 & \textbf{15} &
{61} & {68} & {76} &
\textbf{69} & {66} & {72} &
{60} & {67} & {73} \\
SMART+FT & Alibaba &
{95} & {45} & {93} & \NA & 22 &
{66} & {74} & {81} &
{63} & {72} & {77} &
{62} & {72} & {76} \\
SMART+Al & Alibaba &
\textbf{96} & {47} & {93} & \NA & 25 &
{66} & {75} & {82} &
{64} & {73} & {78} &
{64} & {73} & {78} \\
Env.+FT & Exp. Studies &
\NA & \NA & \NA & {3.6} & \NA &
{66} & {75} & {82} &
{64} & {73} & {79} &
{66} & \textbf{74} & \textbf{81} \\
Env.+Al & Exp. Studies &
\NA & \NA & \NA & {3.5} & \NA &
{67} & \textbf{76} & \textbf{83} &
{65} & {74} & \textbf{80} &
\textbf{67} & {75} & \textbf{82} \\
\bottomrule
\end{tabularx}
\end{table*}

\begin{table*}[!hbt]
\centering
\caption{SSD Operational Data Analysis (\textbf{Fleet Level}). [Abbreviations as above]}
\label{tab:eval-table2}
\renewcommand{\arraystretch}{0.9}
\small
\setlength{\tabcolsep}{4.5pt}
\begin{tabularx}{\textwidth}{X X c c c c c c c c c c c c c}
\toprule
\textbf{Dataset Type} & \textbf{Dataset Used} &
\multicolumn{4}{c}{\textbf{Predictive}} &
\multicolumn{3}{c}{\textbf{Descriptive}} &
\multicolumn{3}{c}{\textbf{Prescriptive}} &
\multicolumn{3}{c}{\textbf{What-if}} \\
\cmidrule(lr){3-6} \cmidrule(lr){7-9} \cmidrule(lr){10-12} \cmidrule(lr){13-15}
& &
\makecell{\textbf{P}\\ $\uparrow$} &
\makecell{\textbf{R}\\ $\uparrow$} &
\makecell{\textbf{A}\\ $\uparrow$} &
\makecell{\textbf{TTF-}\\ \textbf{MSE} $\downarrow$} &
\makecell{\textbf{B4}\\ $\uparrow$} &
\makecell{\textbf{RL}\\ $\uparrow$} &
\makecell{\textbf{FiP}\\ $\uparrow$} &
\makecell{\textbf{B4}\\ $\uparrow$} &
\makecell{\textbf{RL}\\ $\uparrow$} &
\makecell{\textbf{FiP}\\ $\uparrow$} &
\makecell{\textbf{B4}\\ $\uparrow$} &
\makecell{\textbf{RL}\\ $\uparrow$} &
\makecell{\textbf{CFV}\\ $\uparrow$} \\
\midrule
SMART & Alibaba &
{82} & 32 & {89} & {30} &
{62} & {69} & {73} &
{62} & {65} & {64} &
55 & {70} & {67} \\
SMART & Google &
{85} & 37 & {91} & 35 &
{63} & {65} & {70} &
{60} & {69} & {63} &
51 & {69} & {65} \\
\textbf{SMART+Workload} & Alibaba &
\textbf{87} & \textbf{41} & \textbf{93} & \textbf{24} &
\textbf{67} & \textbf{73} & \textbf{77} &
\textbf{64} & \textbf{71} & \textbf{74} &
\textbf{61} & \textbf{70} & \textbf{71} \\
\bottomrule
\end{tabularx}
\end{table*}
We evaluate the operational implications of KORAL’s design choices, including the intermediate representation, the Data Knowledge Graph, the Literature Knowledge Graph, and modality fusion. Device level results are reported in Table~I and fleet level results are reported in Table~II, while Figure~\ref{fig:compariosn} compares reliability prediction against learned baselines on Alibaba SMART windows and Figure~\ref{fig:ablation} quantifies the impact of major components via ablation.

\subsubsection{Reliability prediction across scale and datasets}
At the device level, KORAL improves failure occurrence precision and accuracy and reduces time to failure error because the intermediate representation retains daily trends, change points, and uncertainty that would otherwise be flattened. Recall increases when workload or environment are added to SMART since the model can distinguish short spikes from sustained drift and relate known stressors to elevated risk. Fleet level prediction is inherently harder because one decision aggregates many devices and operating regimes, so precision and recall are lower than at the device level, yet adding workload to SMART at fleet scale still improves reliability metrics and lowers time to failure error, indicating that the representation and aggregation preserve useful signal at scale.

Generalization is evaluated across Alibaba and Google cohorts. SMART only prediction transfers without retuning, which suggests that the intermediate representation plus Data Knowledge Graph (KG) yield stable features despite vendor naming differences and noise. Across cohorts, adding workload or environmental context consistently raises recall with minimal precision change, and it also improves descriptive, prescriptive, and what-if metrics, showing that richer context helps while robustness is maintained across sources. 

\subsubsection{Impact of modality fusion}
Adding workload to SMART increases recall and improves the faithfulness of natural language outputs because workload burstiness and queue depth explain shifts in error counts and latency. For example, under SQL Services workload, daytime concurrency spikes and nightly index rebuild behavior create queue depth surges and higher 95th percentile latency, and with workload context the model interprets brief error bursts as pressure episodes rather than immediate failure.

Adding environmental context improves recall and tail latency estimation because the intermediate representation encodes temperature and humidity with units and links them to attribute frames. During a chiller switchover, inlet temperature rises from 24$^\circ$C to 31$^\circ$C for about four hours with relative humidity near 60\%, and the graph links this excursion to higher latency and correctable error spikes.

Using SMART, workload, and environment together yields the best predictive performance and the strongest grounding, achieving precision of 96\% and recall of 58\% at the device level while also reducing time to failure error, since the generator can cite operational load and ambient conditions for each claim instead of relying on implicit assumptions.

\subsubsection{Grounded natural language analysis and the roles of the two knowledge graphs}
The Data KG is the primary source of factual content for both prediction and descriptive summaries because each claim can be traced to an intermediate representation frame with units, window scope, and coverage, which improves BLEU and ROUGE as well as Faithfulness Precision. This structure rewards grounded statements that connect recent trends and co occurring signals, for example increases in uncorrectable errors over the last days paired with concurrent shifts in controller level behavior, because each atomic claim can be checked against the corresponding attribute frame.

Retrieval from the Literature KG contributes most strongly to prescriptive and what-if quality because it supplies mechanism level evidence and constraints that justify actions and counterfactual directions. On Alibaba SMART at the device level, prescriptive Faithfulness Precision improves from 74 with SMART alone to 80 with SMART plus environment, and Counterfactual Validity increases from 67 to 79 in Table~I. At the fleet level, moving from SMART to SMART plus workload raises prescriptive Faithfulness Precision from 64 to 74 and Counterfactual Validity from 67 to 71 in Table~II. These gains occur when graph queries retrieve studies on thermal sensitivity, humidity effects, and vibration coupling, enabling KORAL to reuse validated mechanisms and thresholds while keeping telemetry grounded in the Data Knowledge Graph.

\begin{figure}[!hbt]
\vspace{-0.2cm}
\centering
\includegraphics[width=\linewidth]{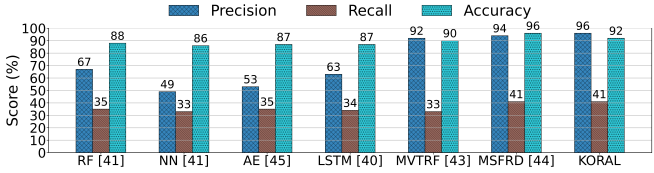}
\caption{Comparison with baselines on Alibaba SMART data.}
\label{fig:compariosn}
\vspace{-0.2cm}
\end{figure}

\subsubsection{Comparison, ablation, and a end to end example}
As shown in Figure~\ref{fig:compariosn}, KORAL surpasses learned baselines on Alibaba SMART data with $30$ day windows by preserving temporal detail and reasoning over multiple modalities through an intermediate representation and two knowledge graphs. Random forest on raw SMART ignores temporal evolution and struggles under drift with weaker recall \cite{b24}, the feed forward neural network on raw SMART is sensitive to class imbalance and drift and needs frequent retuning while lacking explicit temporal structure and operator facing health cues \cite{b24}, and the LSTM needs long well labeled histories, yields opaque scores for rare failures, and still operates as a single view learner that trails MVTRF and MSFRD \cite{b42,mvtrf,msfrd}. Even MVTRF and MSFRD require feature curation and periodic retuning as distributions shift \cite{mvtrf,msfrd}. KORAL preserves daily trends and change points in a rule based intermediate representation, records units and coverage in the Data KG, fuses SMART with workload and environment without retraining, and grounds recommendations and counterfactuals using retrieved literature. This modality aware pipeline improves precision and accuracy, boosts recall when context is added, and reduces error for time to failure and tail latency. By combining paired Data and Literature knowledge graphs with explicit provenance and citations, KORAL supports predictive, descriptive, prescriptive, and what-if analysis across modalities without retraining.

\begin{figure}[!hbt]
\vspace{-0.2cm}
\centering
\includegraphics[width=\linewidth]{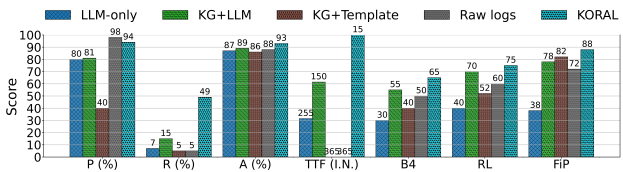}
\caption{KORAL Ablation: end-to-end impact of IR and knowledge grounding on failure prediction, TTF estimation, and analysis quality (Abbreviations as in Table I).}
\label{fig:ablation}
\end{figure}
We ablate KORAL on Alibaba SMART plus Workload with a thirty day window as shown in Figure~\ref{fig:ablation}. Time to failure (TTF) is normalized and inverted to align with the axis. Text metrics are reported as a combined analysis score by sampling prompts from descriptive, prescriptive, and what-if tasks and averaging over windows. Across ablations, partial pipelines tend to be overly conservative: precision can remain high, but recall collapses, TTF error grows, and the generated analyses become less consistent. In particular, KG plus LLM mainly improves the readability and coverage of generated summaries, while KG plus Template increases faithfulness oriented scores at the cost of weaker failure sensitivity, and raw logs behaves like hard thresholding that misses many failures.

In contrast, KORAL is consistently strongest across both predictive and generative axes. With all components enabled, it maintains high precision while substantially improving recall, achieves the best overall accuracy, and yields the lowest TTF error, indicating earlier and more reliable failure anticipation. It also produces the most informative and faithful analyses, leading on BLEU 4, ROUGE L, and Faithfulness Precision, which reflects better instruction following and background preservation in the generated reports. These gains come from the synergy of the intermediate representation with the Data KG and Literature KG: the IR captures temporal structure and change points, the Data KG anchors numeric claims and cohort context, and the Literature KG adds mechanism level grounding that strengthens prescriptive and counterfactual reasoning.

\textbf{Example.}
Consider drive D42 at site Alpha in a window where the summary records a temperature rise from 40$^\circ$C to 58$^\circ$C, an increase in write share from about 60\% to 80\%, a higher program fail rate, step increases in write amplification aligned with workload changes, and missing airflow on two days. KORAL encodes these as typed frames and graph nodes linked to the device and window, then retrieves literature about thermal stress under write heavy workloads. The resulting output can explain the observed latency and error behavior using data grounded evidence, propose operational actions such as reducing write burst intensity or improving airflow when retrieved evidence supports it, and answer what-if queries such as the expected change if queue depth is capped or ambient temperature is reduced, while keeping each claim traceable to either the Data KG or retrieved literature context.

\section{Related Work}
\label{sec:related_work}
Prior work on SSD operational analysis and system logs are diverse, covering environmental impacts, SMART-based reliability and failure prediction, and broader observability and performance diagnosis. 
% We structure this section along these axes and then summarize recent LLM-based approaches for operational reasoning over system logs.

\textbf{Environmental impact and operating conditions.}
A line of work shows that external conditions materially affect SSD behavior and reliability. Experimental studies examine how temperature, humidity, and vibration exposures influence latency, performance variability, and degradation trends~\cite{b1,b2}. Complementary efforts focus on thermal effects inside the device, including predicting throttling behavior~\cite{b16} and improving NAND reliability by explicitly exploiting temperature awareness and self-recovery properties~\cite{b18}. Beyond immediate reliability, sustainability-oriented studies highlight the broader environmental footprint of storage devices, including embodied carbon considerations~\cite{b8}. Most recently, LLMs have been explored as a modeling layer to connect environmental factors with SSD outcomes and summarize multi-factor impacts from heterogeneous evidence~\cite{ssd_llm_akewar}. These works motivate KORAL’s explicit treatment of environment as first-class operational context, but they typically do not unify environmental signals with workload, SMART telemetry, and literature knowledge in one structured reasoning pipeline.

\textbf{SMART telemetry and SSD failure prediction.}
A substantial literature uses SMART attributes and fleet telemetry to characterize failure modes and build predictive models. Large-scale field studies report failure symptoms, root causes, and operational failure patterns at scale~\cite{b11,b12,b23,b24,b25,b35}, while postmortem analyses of SSD-related incidents highlight practical lessons from real deployments~\cite{b14}. Surveys summarize the broader design space of lifecycle prediction and modeling choices~\cite{b10}. On the modeling side, feature selection and supervised learning remain common for failure prediction in large deployments~\cite{b30}, and more recent approaches improve adaptability and interpretability of predictors~\cite{b28,b42,mvtrf,msfrd}. While these methods can achieve strong predictive accuracy, they often output scores/labels with limited operator-facing explanation, and they generally do not connect predictions to structured evidence from documentation and prior studies in a way that supports \emph{descriptive, prescriptive, and what-if} operational queries.

\textbf{LLMs for operational reasoning and system logs.}
Recent systems increasingly treat LLMs as an ``ops copilot'' that can read noisy operational artifacts, including logs, traces, and configuration metadata, and produce summaries, anomaly hypotheses, and remediation suggestions~\cite{llm1,llm2,log_llm}. A parallel thread explores tool-augmented (often agentic) incident workflows, where the LLM is coupled with log search, dashboards, and code/context inspection to support root-cause analysis and incident triage, including multi-step and multi-agent variants for verification and action planning~\cite{llm3,llm4}. Beyond diagnosis, LLMs have also been used to navigate large tuning/control spaces: e.g., guiding HPC I/O optimization and configuration exploration, assisting database/LSM tuning, and pairing natural-language goal specification with policy learning for scheduling and control in heterogeneous environments~\cite{b34,b33,llm6,llm12, DBLP:conf/ipps/EgersdoerferSBB25}. Within storage management, initial work applies LLMs to high-level tiering and parameter selection for hybrid SSD deployments~\cite{llm_ssd}, while recent efforts study LLM+RAG pipelines to connect environmental factors with SSD performance and reliability outcomes from heterogeneous evidence~\cite{ssd_llm_akewar}. However, log-centric LLM pipelines can still over-generalize when signals are incomplete or when domain constraints are implicit, making it hard to guarantee that explanations are evidence-backed. KORAL addresses this gap by grounding generation in two explicit KGs and performing structured retrieval over them, enabling citations-to-evidence style answers that unify telemetry, environment, workload context, and prior studies across descriptive, predictive, prescriptive, and what-if SSD queries.

\section{Limitations}
\label{sec:con_fut}
KORAL provides exploratory analysis over scattered telemetry and shows promising results, yet important limitations remain. Our evaluation draws on controlled environmental studies and production telemetry that do not cover every deployment setting, and ground truth for rare failures and performance deltas is sparse and imbalanced. Operator query logs are limited, so some what-if and descriptive tasks are under represented. The prescriptions have not been tested in production, so some recommendations may be incorrect. We did not evaluate open source LLMs and relied on a proprietary model, which can introduce training data bias and leaves the performance of KORAL with open source models unknown.
% KORAL is a novel framework that shows promising results on scattered telemetry and provides exploratory analysis, but important limitations remain. Our evaluation relies on controlled environmental studies and production telemetry that do not cover every deployment setting, and ground truth for rare failures and performance deltas is sparse and imbalanced. Operator query logs are scarce, so not all what-if and descriptive tasks are represented. The prescriptions produced by the framework have not been tested in production, so some recommendations may be incorrect. We did not evaluate open source LLMs and relied on a proprietary model, which may introduce training data bias, and the performance of KORAL with open source models is not yet known. We have not explored alternative prompt strategies also the intermediate representation uses a manually curated rule base, and alternative rule bases and ablations are not yet evaluated.
\section{Conclusion}
\label{sec:conclusion}
This study presented KORAL, a knowledge driven framework that combines LLMs with two KGs to support SSD operational analysis from fragmented evidence. We address RQ1 by extracting and organizing evidence backed facts from prior studies into a Literature KG, so established results remain reusable during operation. We address RQ2 by converting production telemetry into a structured Data KG that preserves temporal and fleet context for summarization and prediction. We address RQ3 by grounding generation in graph retrieval and domain constraints, which improves faithfulness and reduces unsupported claims. We address RQ4 by enabling grounded descriptive, predictive, prescriptive, and what-if analysis in a single workflow, with evaluations showing that KORAL produces actionable insights with transparent supporting evidence. Although we focus on SSDs, the architecture is not device specific, and it can be adapted to other devices such as GPUs by swapping the ontology and telemetry sources.

Next, we will broaden coverage by incorporating in rack sensors, additional vendors and device models, and continuous production telemetry. We will expand expert review, adopt active labeling, and track long horizon cohorts to improve coverage and reduce bias. We will compare multiple open source language models with the current model and apply domain specific fine tuning, instruction tuning, and calibration to improve faithfulness and robustness. Finally, we will enlarge the query set with operator authored tasks, collect real operator queries, and evaluate prescriptive and what-if tasks more systematically.
\section*{Acknowledgments}
This work was supported by the following NSF grants: CSR-2402328, CAREER-2338457, CSR-2406069, CSR-2323100, HRD-2225201, 2529283 and 2242700, as well as by the DOE Office of Science User Facility, supported by the Office of Science of the U.S. Department of Energy under Contract No. DE-AC02-05CH11231, using NERSC award DDR-ERCAP0035598. This research was also supported by the U.S. Department of Energy, Office of Science, Advanced Scientific Computing Research, through the SciDAC-RAPIDS2 institute under Contract DE-AC02-06CH11357 and Laboratory Directed Research and Development (LDRD) funding from Argonne National Laboratory under Contract DE-AC02-06CH11357. We thank Nityam Sharadkumar Bhimani (Vertex Pharmaceuticals) for inspiring the core idea of leveraging knowledge graphs to guide and constrain LLM reasoning. 
% We also sincerely thank the anonymous reviewers for their valuable feedback. 

\bibliographystyle{IEEEtran}
\bibliography{references}

\end{document}